**Title:**

Revealing Nanoscale Molecular Organization in Liquid Crystals via Cryogenic Atom Probe Tomography

**Authors:**

Kuan Meng[1], Kang'an Wang[2], Sebastian Eich[1], Pierre Nacke[3], Johanna R. Bruckner[3], Patrick Stender[1], Frank Giesselmann[3], Guido Schmitz[1]

**Affiliations:**

1) University of Stuttgart, Institute of Materials Science, Germany
2) University of California, Berkeley, Department of Materials Science and Engineering, USA
3) University of Stuttgart, Institute of Physical Chemistry, Germany

**Corresponding Author Email Address:**

Kuan.Meng@mp.imw.uni-stuttgart.de

**Abstract:**

While liquid crystals (LCs) have been extensively studied, obtaining a comprehensive nanoscale picture of their molecular organization remains challenging, as conventional techniques face an intrinsic trade-off between spatial and chemical resolution. Here, cryogenic atom probe tomography (cryo-APT) is introduced as a new analytical approach for LC materials, using 4'-Pentyl-4-cyanobiphenyl (5CB) and 4'-Octyl-4-cyanobiphenyl (8CB) as representative model compounds. This was enabled by a tailored cryogenic focused ion beam (cryo-FIB) protocol optimized for small organic molecules. The method enables controlled field evaporation of both intact molecules and diagnostic fragments, achieving over 90% molecular retention while preserving four characteristic dissociation patterns. By spatially correlating these fragmentation profiles with the local electric field derived from the tip geometry, we reveal field-directed dissociation pathways of CB molecules. In parallel, the distribution of intact molecular ions enables nanoscale visualization of material structure: we resolve homogeneous mixing of 5CB and 8CB in the nematic phase and directly observe the sub-nanometer crystalline layering in a supercooled 8CB sample, with contrast to the surrounding amorphous matrix suggesting the presence of a solid-liquid interface. This work establishes cryo-APT as a new powerful analytical platform for LC research and reveals its broad potential for application in soft matter systems.

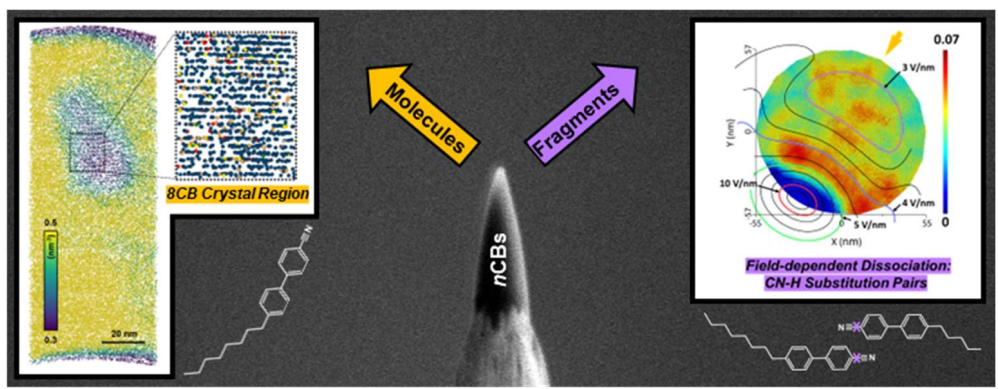

1. **Introduction**

Liquid crystals (LCs) are mesophase materials that combine the fluidity of liquids with the molecular long-range order of solids.[1] This duality underpins their transformative role in modern technologies, most notably flat-panel displays,[2,3] and their ubiquity in biological systems,[4] including lipids,[5] carbohydrates,[6] proteins,[7] and nucleic acids.[8] As a result, LC research has become an interdisciplinary field, bridging physics, chemistry, biology, and technology. Yet, at the core of this vast field lies a fundamental question: how anisometric organic molecules align and organize in response to external stimuli.

Unraveling molecular responses in LCs requires advanced structural characterization across multiple length scales.[9] Optical microscopy[10] captures the director structure in bulk samples but lacks the resolution to probe molecular-level organization. Scattering techniques, such as neutron and X-ray scattering,[11,12] provide valuable insights into molecular ordering in LC phases. Calculated electron density maps enable visualization of complex supramolecular architectures.[13,14] Spectroscopic methods, including nuclear magnetic resonance (NMR),[15] electron paramagnetic resonance (EPR),[16] Raman and infrared spectroscopy,[17,18] quantify order parameters for specific molecular segments. However, all these techniques rely inherently on ensemble averaging, which obscures microscopic details. Imaging techniques, such as transmission electron microscopy (TEM) and cryo-TEM[19,20] offer excellent spatial resolution but struggle with chemical contrast of light elements, particularly in the mixed systems where components could only differ in aliphatic length.[21] Freeze fracture TEM reveals fine 2D structure features[22–24] yet sacrifices 3D spatial information in bulk.

While these methods have significantly advanced the understanding of LCs, a critical limitation remains: existing tools excel either in spatial resolution or in chemical contrast but rarely achieve both simultaneously. Bridging this gap could offer novel insights into persistent challenges in LC research.

Atom probe tomography (APT) emerges as promising tool to overcome these limitations, combining sub-nanometer spatial resolution,[25] sub-unitary mass spectral resolution (<1 amu),[26] and high chemical sensitivity (<100 ppm). This 3D chemical mapping technique could provide novel structural insights to enhance the structural elucidation of LC systems. In APT, a tip specimen with a sub-micrometer radius of curvature is cooled to cryogenic temperatures and subjected to high bias to create a strong local electric field. Entities (atomic, molecular or clustered ions) are then field evaporated layer by layer, triggered by voltage or laser pulses, and captured on a 2D positional and time-sensitive detector. The ions' species are identified through time-of-flight (ToF) analysis, allowing for the reconstruction of the sample's 3D chemical distribution.

While APT has shown great progress in nanoscale chemical analysis, its application to LC systems remains underexplored. The necessity for a precise tip geometry has remained a key bottleneck, limiting APT applications to rigid or easily shaped materials and excluding many soft organic molecules like LCs. As a result, current research often relies on deposition onto pre-prepared needle-shaped substrates,[26–28] which naturally emphasizes the adsorption behavior, with relatively little focus on the bulk properties. Moreover, the sample morphology of the deposition layer is microscopically uncontrollable, which could influence the bonding and fragmentation behaviour.[28] The very recent advancements in cryo-focused ion beam (cryo-FIB) technology[29–32] have made it possible to investigate the bulk behavior of frozen liquids, offering valuable insights for aqueous-based systems.[33–36] However, its application to non-aqueous liquids, such as organic molecular systems,[37] remains in its infancy. Therefore, establishing an analytical protocol tailored for LC systems would not only advance the understanding of organic molecules but also extend the capabilities of cryo-FIB and APT techniques.

In this study, we applied cryo-APT for the first time to investigate LCs. We selected two well characterized LC compounds: 4′-Pentyl-4-cyanobiphenyl (5CB), the first commercially viable LC for display technologies[38] and 4′-Octyl-4-cyanobiphenyl (8CB), the first nCB homolog to display a smectic phase.[39] Both components consist solely of light elements—carbon, hydrogen, and nitrogen—while their only difference is three additional methylene (-$CH_2$) units in the 8CB alkyl chain. First, we developed an experimental protocol for LC tip preparation originating from fluid state using cryo-FIB technique. Next, we analyzed the two compounds individually to assess their stability during preparation and field evaporation, identifying the molecular and fragment signals while examining their spatial distributions to understand the fragmentation process. Finally, we investigated the molecular distribution of the two components in the mixture and resolved the crystalline structure of 8CB down to lattice spacing details.

## 2. Results and Discussion
### 2.1 Tip Preparation

Organic LCs, unlike aqueous systems, present overlooked challenges for cryo-FIB shaping from the fluid state. Their low thermal conductivity[40,41] often leads to local overheating and deformation (Fig. S1), while high electron and ion beam sensitivity could induce bond cleavage and alter composition by ion mixing.[42,43] Most critically, gallium implantation forms a dense modified surface layer that strongly inhibits further sputtering and significantly slows down the shaping process.[44] These factors collectively demand a protocol that preserves geometric precision and chemical fidelity within a feasible time effort (< 7 h). To address this, we adapted the wire-based approach[30] for organic LC tip fabrication (Fig. 1).

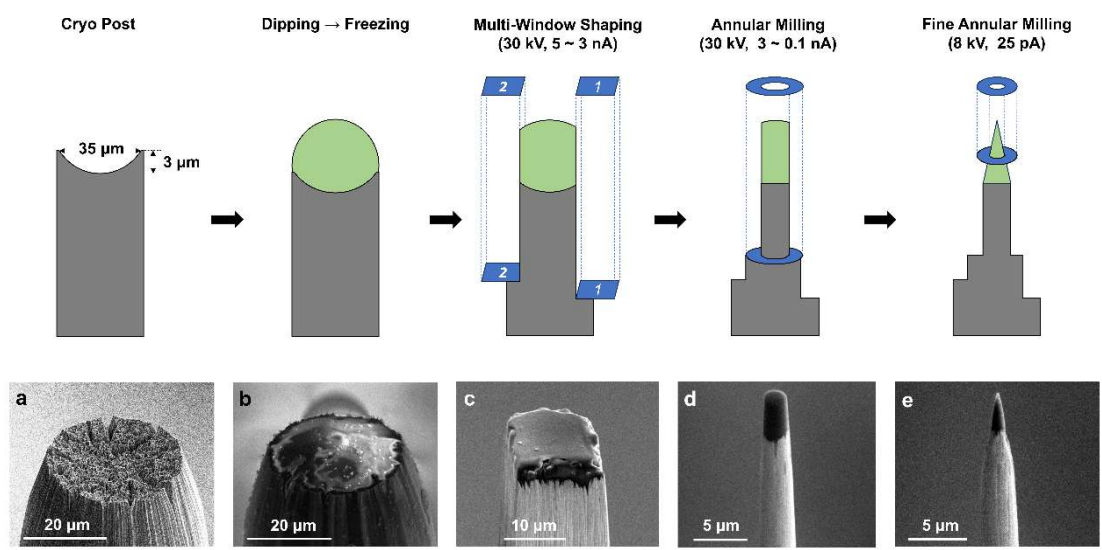

*Figure 1. Schematic of tip preparation procedure for liquid crystal and corresponding SEM pictures at each stage.*

To improve tip preparation efficiency, we minimized the initial deposition volume. This was achieved by cryogenically fracturing a 50 µm tungsten wire under tensile stress, resulting in a shallow concave well (Fig. 1a) suitable for LC deposition. The droplet was applied with a glass Pasteur pipette prior to freezing, yielding a uniform coating under 10 µm (Fig. 1b). The sample was quench-frozen in liquid nitrogen at a rate of ~70 K/min, well above the crystallization threshold of 5CB (~5 K/min),[45] to preserve the native molecular order. It was subsequently transferred for freeze-etching, then mounted in a cryo dual-beam FIB/SEM system for shaping.

Due to the extreme beam sensitivity of LC samples, electron (5 kV, 50 pA) and ion (<50 pA) imaging were restricted to low intensities to avoid radiation damage.[46] Importantly, no imaging was performed at higher ion currents above 0.3 nA during milling, as even brief exposure irreversibly damaged the LC sample (Fig. S1a-b). Initial shaping began with annular milling at 5 nA and 30 kV to reduce droplet size, but unfortunately, circular heating induced hollowing near the center (Fig. S1c–d), preventing further progress. To mitigate this, we employed a multi-window shaping strategy,[37] using a series of tangentially arranged rectangular milling patterns to confine beam exposure to only one side at a time. Once the lateral size was reduced below 20 µm (Fig. 1c), conventional annular milling resumed with stepwise reduction in ion current to refine tip geometry. At least 10 μm of both LC and tungsten was removed to eliminate secondary protrusions near the apex. Final shaping was performed under 8 kV and 25 pA below a 3 μm diameter, terminating at an inner tip diameter of 200 nm.

By employing this protocol, we achieved precise geometric control of LC tips: all nine specimens exhibited a uniform shaft angle (~20°) and a curvature radius exceeding 100 nm (Fig. S2). The entire fabrication process of a sample was completed within 7 hours, marking a substantial improvement over previous study(> 10 h)[37]. Remarkably, the method initiates from the highly deformable fluid state, underscoring its robustness and broad adaptability for future applications in LCs (Fig. S2) and organic systems.

**2.2 Field Evaporation Behaviors of *nCB* Molecules**

Seeing nanoscale structure in LCs *via* APT begins not with resolution, but with interpretation. Yet in soft matters, field-driven dissociation and complex fragmentation patterns challenge the interpretation of even the most carefully collected spectra. Specifically, three challenges arise. Firstly, identical molecules often yield distinct mass spectra under different field conditions,[28,30] raising the critical question of which spectra can be reliably used for structural interpretation. Secondly, even within a single spectrum, the sheer number of fragmentation possibilities often lead to ambiguity in assigning molecular or fragment identities.[34] Thirdly, low-abundance signals, such as fragment peaks, are often dismissed as noise. However, APT's high chemical sensitivity invites a deeper consideration: could we uncover their hidden chemical insights?

To ensure reliability, each system (pure 5CB, 8CB, and their binary mixture) was measured at least three times under consistent conditions. Analyses focused exclusively on regions where evaporation was uniform and only reproducible spectra were considered (Fig. S3-4). A representative dataset from each system was then selected for detailed examination. Apart from reproducibility, the pure 5CB measurement (Fig. 2) with over 17 million events and 354 million assigned atoms, provided a reliable statistical analysis. Beyond quantity, the most compelling evidence for the quality of the 5CB spectrum lies not in what was detected, but in what was missing.

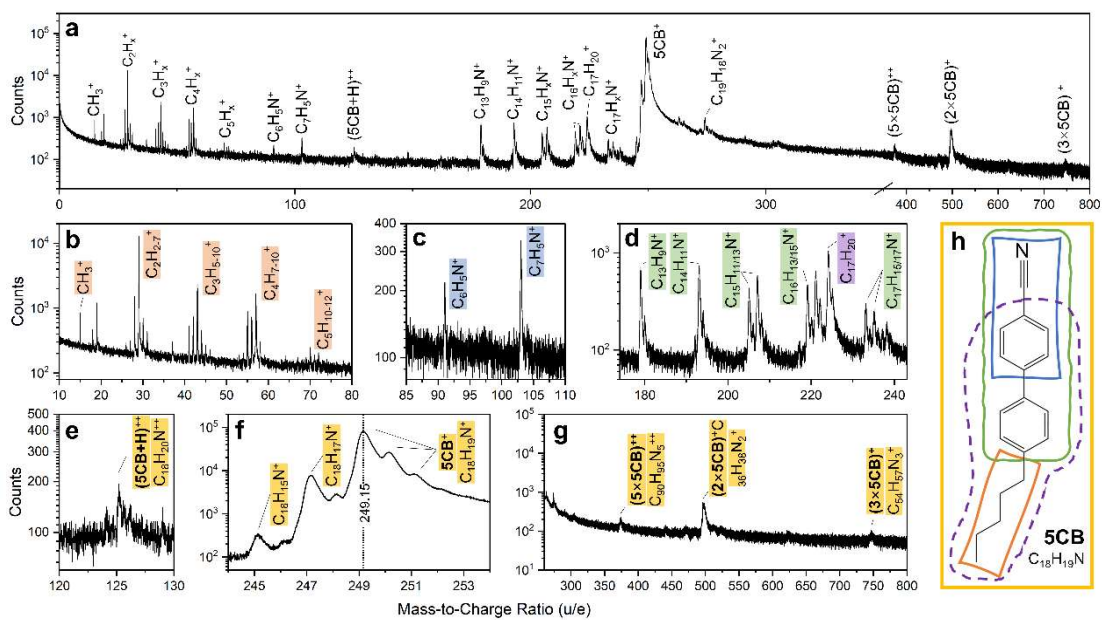

*Figure 2. Mass spectrum of pure 5CB. a) Logarithmic view from 0 to 800 u, with 0-350 u expanded for clarity. b-d) Fragments signals, including alkyl fragments (orange), monophenyl derivatives(blue), cyanobiphenyl residues (green), and CN-H substitution pairs (purple). e-g) Full molecule signals labeled in yellow. h) Schematic of fragmentation behavior of 5CB.*

First, water-related peaks ($OH_2^+$, $OH_3^+$) near 18 and 19 u accounted for less than 0.12% of events (Table S2). This negligible amount aligns with the hydrophobic nature of 5CB, indicating the purity of the sample and minimal background interference. Second, the absence of $Ga^+$ signal near $m/z$ = 69 u (Fig. 1b) indicates that the presented preparation strategy successfully prevented gallium implantation, thereby preserving the chemical integrity of the LCs. Notably, the spectrum exhibited a complete absence of atomic signals ($H^+$, $C^+$, $N^+$). This rare but revealing case underscores the complexity of field evaporation behavior in organics and consequently complicates spectral interpretation.

To mitigate the ambiguity in peak assignment, we employed a bin size of 0.01 u for accurate peak positioning and spectrum fitting for stoichiometry calculation (Fig. S7). A full breakdown of three measured systems is provided in table S2-4. Overall, the 5CB spectrum was dominated by intact molecular signals, with only minimal fragmentation (Fig. 2a).

Approximately 92.22 % molecules evaporated intact or with minor $H_2$ loss (Table S5), reflecting the stability of 5CB under the applied electric field. The largest contribution stemmed from the isotope peak cluster of singly charged monomers $C_{18}H_{19}N^+$ and its dehydrogenated variants $C_{18}H_{17}N^+$ and $C_{18}H_{15}N^+$ (Fig. 2f), followed by singly charged dimers $C_{36}H_{38}N_2^+$, trimers $C_{54}H_{57}N_3^+$, doubly-charged ions $C_{18}H_{20}N^{++}$ and oligomers $C_{90}H_{95}N_5^{++}$ (Fig. 2e, g). This predominant molecular evaporation suggests that the weak intermolecular van der Waals bonds are easily overcome under the applied field strength, while the intramolecular covalent bonds remain stable.

Despite the dominance of intact molecules, the presence of fragment species (here < 8%) holds scientific value for understanding potential dissociation pathways. For clarity, fragment signals were grouped into four main categories: alkyl-chain fragments, monophenyl derivatives, cyanobiphenyl residues and cyano-hydrogen (CN-H) substitution pairs. These were consistently color-coded as orange, blue, green and purple, respectively (Fig. 2h).

Alkyl chain fragments from aliphatic $C_1$ to $C_5$ peak groups (with varying amount of hydrogen) match the alkane chain length of 5CB (Fig. 2b). For monophenyl derivatives, the cyanophenyl fragment $C_6H_5CN^+$ around $m/z$ = 103 u and its derivative $C_6H_5N^+$ about $m/z$ = 91 u were identified (Fig. 2c). As cyanobiphenyl residues, five peak groups from $C_{17}$ to $C_{13}$ appear in intervals of 14 u, each representing the loss of a further methylene fragment ($-CH_2$) from the alkyl chain. In the CN-H substitute pairs, peaks close to $m/z$ = 249 ± 25 u were identified and indicate two effects. One pathway involves hydrogen substitution of the cyano group, yielding the $C_{17}H_{20}^+$ ion near $m/z$ = 224 u. In another, the cleaved CN group replaces a hydrogen atom in a neighboring 5CB molecule, forming $C_{19}H_{18}N_2^+$ near $m/z$ = 274 u. The presence of $C_{19}H_{18}N_2^+$ ions and other peaks beyond the molecular positions suggests that field evaporation does not involve exclusively fragmentation. Instead, recombination of fragments prior to detection should also be considered.[28,37,47,48] To validate our peak assignments, we calculated the chemical stoichiometry only from the fragment signals (Table S6). Despite detector efficiency limits (~50 %), the average deviation from theoretical elemental ratios remained below 4%, demonstrating sufficient accuracy of the calibration-free measurement and reliable peak interpretation.

To assess whether the evaporation pattern seen in 5CB extends across the CB homologues, we examined 8CB under identical conditions. As a structural homologue, 8CB exhibits a mass spectrum closely resembling that of 5CB (compare Figs. 3a, b). Minor differences arise from the longer alkyl chain in 8CB, leading to extended alkyl fragments $C_{6-8}H_x^+$, longer cyanobiphenyl residues $C_{18-20}H_xN^+$, and a systematic shift of molecular peaks to higher $m/z$. In particular, the fraction of intact molecules in 8CB (91.44 %) closely matches that in 5CB (92.22 %), suggesting that fragmentation behavior remains consistent despite varying chain lengths. Furthermore, the spectrum of the mixed system (Fig. 3c) appears as a direct superposition of the two pure components.

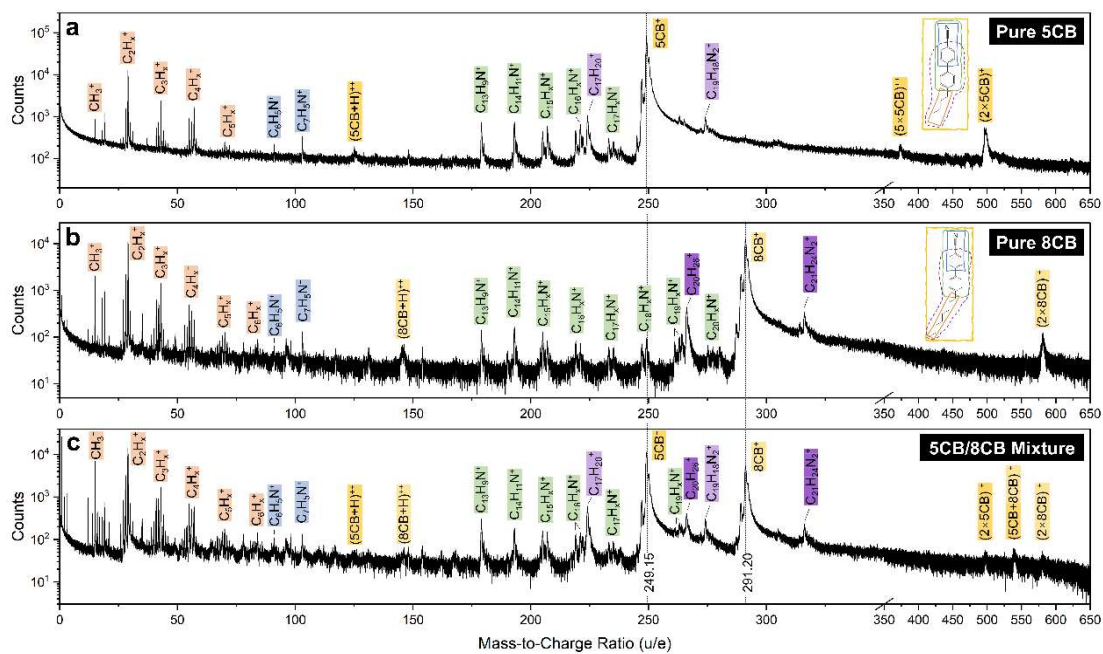

*Figure 3. Mass spectra in direct comparison. a) 5CB. b) 8CB. c) Mixture of $5CB_{(0.55)}8CB_{(0.45)}$, where the molar ratio of 5CB is approximately 55%.*

This spectral consistency confirms both the stability of our field evaporation conditions and the accuracy of peak interpretation. Furthermore, the reproducible detection of low-abundance fragment signals highlights APT's high chemical sensitivity. Rather than noise, these minor signals carry intrinsic chemical insights, as evidenced by two consistent observations across all systems: systematic dehydrogenation and the formation of molecular dimers.

A characteristic signature of dehydrogenation emerged from the cyanobiphenyl residue series. In both 5CB and 8CB, the cyanobiphenyl residues show 14 u regular spacing due to progressive loss of methylene units (-CH$_2$). But only fragments with at least 15 carbon atoms exhibit paired peaks separated by 2 u, characteristic of H$_2$ loss. In contrast, the C$_{13}$ and C$_{14}$ fragments appear as single peaks without this splitting (Fig. 2d, Fig. 3b-c). Considering the C$_{13}$ peak C$_{13}$H$_9$N$^+$ corresponds to the alkyl-free cyanobiphenyl core H$_5$C$_6$C$_6$H$_4$CN$^+$, the appearance of dehydrogenation peaks only in heavier fragments (above C$_{15}$) implies that hydrogen abstraction requires the presence of at least two methylene units along the alkyl chain. Dehydrogenated molecular species (Fig. 2f, Fig. 3b-c) further support this trend. Systematic dehydrogenation has been well documented for saturated hydrocarbons in field desorption and ionization mass spectrometry.[49–52] These findings are particularly relevant to APT, given the comparable evaporation forms and electric field strengths.[53] Species of the form [M-nH$_2$]$^+$ (n≤3), where M denotes the intact molecular ion, have been previously reported.[49] In our data, the [M-H$_2$]$^+$/M$^+$ ratio reaches 8.5% and increases to 16.8 % in 8CB, in agreement with Heine and Geddes's finding that longer alkyl chains promote dehydrogenation.[52]

Another notable discovery is the detection of dimer species in all systems. In particular, a peak near *m/z* = 538 u in the mixed system represents a combined 5CB and 8CB dimer, confirming even the existence of cross-species dimers (Fig. 3c). While dimerization could in principle occur during field evaporation, its consistent presence across all systems and its absence in prior APT studies of n-tetradecane,[37] suggest a structure-specific origin. In fact, cyanobiphenyl dimers have long been established in condensed phases,[54–57] where strong dipolar interactions between terminal cyano groups favor antiparallel configuration. The remarkable stability of such weakly bound aggregates under intense field evaporation not only validates their intrinsic nature but also demonstrates APT's capacity to preserve and reveal intermolecular architectures in soft matter. These same interactions may also support the formation of higher-order aggregates, such as trimers and oligomers, particularly in 5CB (Fig. 2g). Together, these findings highlight APT's potential to investigate molecular interactions in LC systems with unprecedented spatial and chemical resolution.

The strong agreement among the APT spectra, the molecule structure and established references confirms the reliability of both sample preparation and signal interpretation. This foundation enables us to explore how these signals are arranged within the specimen, offering new perspectives and scientific insights in LC research.

**2.3 From Fragmentation Profiles to Molecular Spatial Organization**

Having identified key molecular and fragment species, we focus on two core questions regarding their spatial distribution: 1) How can fragment patterns reflect the field-driven dissociation mechanisms of nCB molecules? 2) To what extent can intact molecular signals reflect structural characteristics across different phases? We begin with the analysis of fragment signals to uncover how environmental conditions, particularly laser illumination and local electrical field, govern molecular dissociation.

To probe this relationship, we examined a representative reconstruction of a 5CB/8CB mixture (5CB, 20 mol%) based on the classical reconstruction model.[58] The signal distribution exhibits pronounced inhomogeneity, which is closely aligned with the direction of laser irradiation (yellow arrows in Fig. 4). On the illuminated side, molecular signals of 5CB and 8CB (blue or turquoise events) accumulate in high density while the shadowed region is dominated by alkyl fragments (orange) with low density (Fig. 4b, "Slice View", "Density"). As measurements progress, the orange region expands while the blue domain recedes, forming an evolving interface over time (Fig. 4a). The anisotropy, confirmed across several independent measurements, is attributed to a laser-induced "shadow effect"[59] amplified by the intrinsically low heat conductivity of the LC.

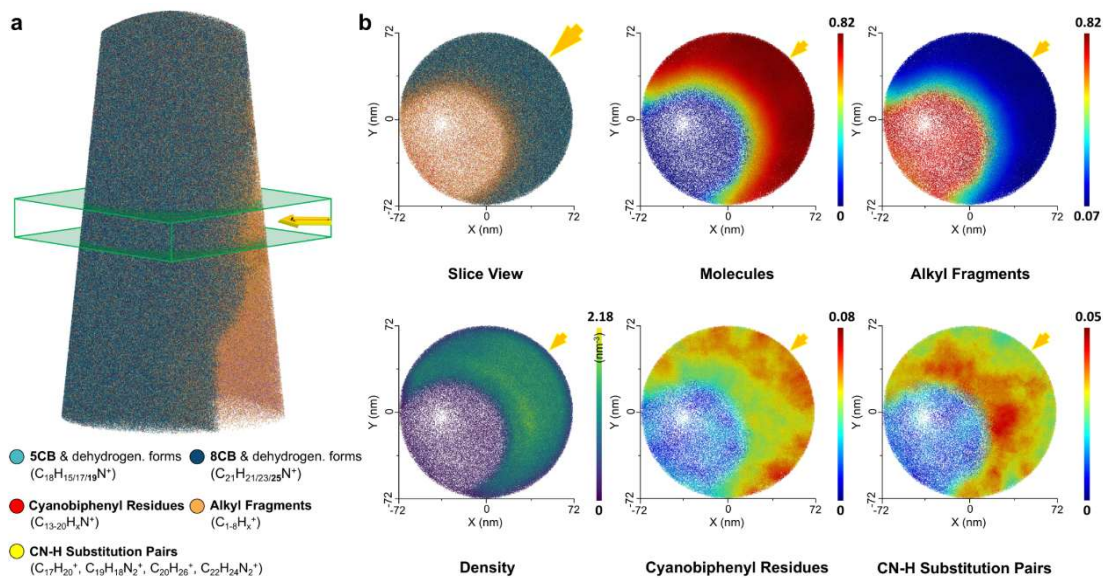

*Figure 4. a) Tip reconstruction of a 5CB$_{(0.2)}$8CB$_{(0.8)}$ mixture, where the molar ratio of 5CB is approximately 20%. b) Fragmentation behavior of cyanobiphenyls analyzed through local density and composition maps of major signals in a 30 nm thick slice, oriented along the direction of laser illumination (yellow arrow).*

The UV laser causes localized heating on the irradiated side, due to unidirectional illumination and its surface-confined absorption depth.[60] While most metallic materials dissipate the heat rapidly, LC materials exhibit slower thermal relaxation, leading to a pronounced temperature gradient across the tip. Under initially homogeneous field conditions, molecules evaporate more frequently from the warmer, illuminated region. This thermal asymmetry distorts the tip shape until the local field strength inhomogeneity compensates the bias in evaporation rate. At steady state, the irradiated side adopts a lower curvature, resulting in a reduced local electric field, while the shadowed side retains a higher curvature and therefore a stronger local field.

This variation in local field strength is critical for interpreting APT data from soft organic materials, as it indicates that field variation does not just change ionization states, but fundamentally alters the detected chemical species. Unlike metallic samples, where field variation mostly changes charge states from the same species,[61,62] nCB exhibits a far more pronounced response: a systematic shift from intact molecule to fragment species. The reduced field allows molecules at the illuminated side to desorb by overcoming only weak van der Waals interactions, resulting in predominant molecular signals (see Fig. 4b, "Molecules"). In contrast, the stronger field at the shadow side induces rupture of covalent bonds, yielding primarily small hydrocarbon signals (see Fig. 4b, "alkyl fragments").

As the evaporation process clearly deforms the tip shape away from the idealized hemispherical assumption used in classical APT reconstructions,[58] this raises a key question: can we reconstruct the actual tip geometry and, based on it, determine the true spatial distribution of chemical species and the corresponding local electric field? The alternative reconstruction is enabled by a key relationship in APT: variations in detection density can be traced back to differences in local magnification,[63] which in turn reflect local curvature of the tip surface. To implement this, we applied the curvature extraction method by Beinke and Schmitz.[64] A rectangular slice containing 2.9 million events (Fig. 4a) was reconstructed to show the real asymmetric tip surface displayed in Fig. 5a. Based on this surface geometry, a density corrected volume reconstruction is obtained, that avoids local magnification artefacts. Furthermore, this geometry allows direct calculation of the local field strength at the tip surface. For this, we applied the mesh-free concept of Rolland *et al.*,[65] to obtain the field as shown in Fig. 5b.

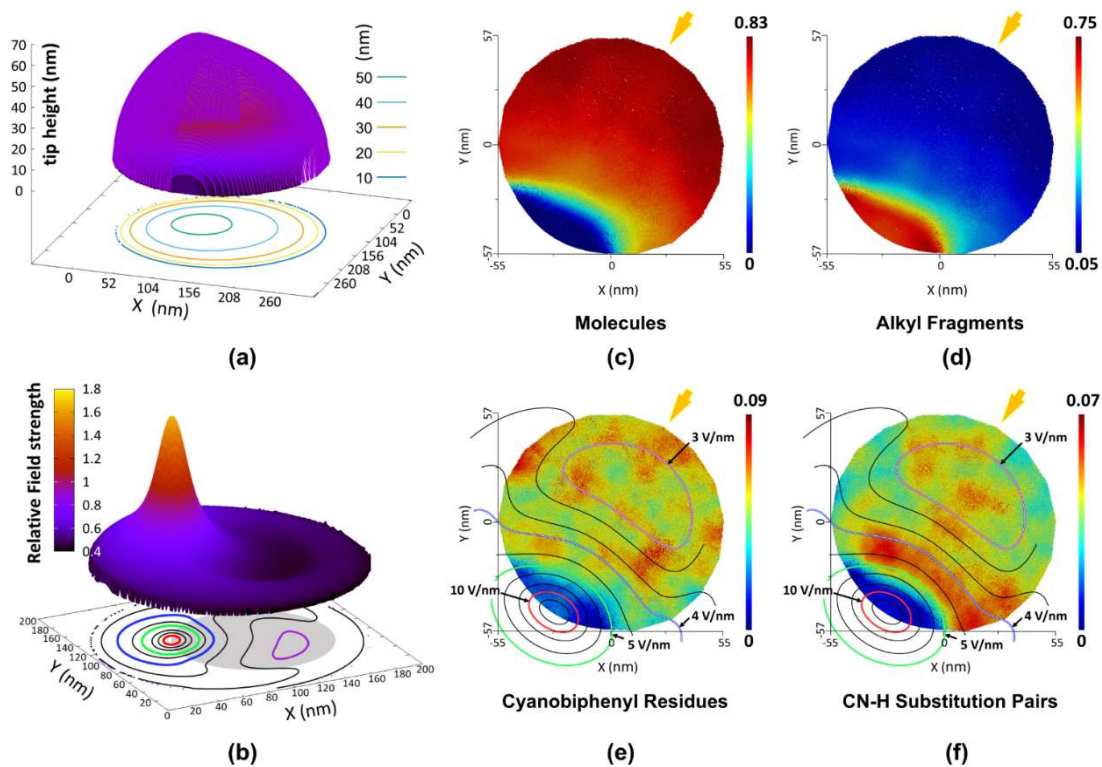

*Figure 5: Field-driven fragmentation behavior revealed via curvature-based reconstruction and spatial field mapping. a) Tip geometry reconstructed from detection density via the curvature extraction method by Beinke and Schmitz.[64] b) Local field strength calculated from the tip shape in (a) by a mesh-free solver of electrostatics according to Rolland et al..[65] The XY plane shows iso-field contours across the entire tip surface. Among them, only the contours within the central shadowed region, which corresponds to the actual reconstructed volume, are used in the overlay with corrected composition maps of major signals. c-d) Lateral distributions of intact molecules and alkyl fragments, respectively. (Note the significant change to the (incorrect) classical reconstructions shown in Fig. 4b). e-f) Iso-field contours overlaid on the corrected maps of cyanobiphenyl residues(e) and CN-H substitution pairs(f), highlighting their preferential field ranges at ~3 V/nm and 4-5 V/nm, respectively.*

Clearly, one sees that the field is rising toward the shadowed end leading to a peak strength before decreasing again when approaching the tip shaft. Figs. 5c and 5d demonstrate the improved (corrected for local magnification) maps of intact molecules and alkyl fragments in comparison to the maps shown in Fig. 4 that were obtained by the classical reconstruction. Finally, the 2D iso-field contours shown in Figs. 5e and 5f allow a quantification of the field strengths at which certain fragments evaporate preferentially.

By overlaying the calculated iso-field contours onto the corrected composition maps, we directly visualize how increasing field strength governs the dissociation pathways of CB molecules (Fig. 5e-f). Cyanobiphenyl residues are localized near the laser-facing edge (~3 V/nm), indicating that initial cleavage occurs along the alkyl chain. CN-H substitution pairs appear in a narrow interface region (4-5 V/nm), which corresponds to bond rupture between the biphenyl core and the cyano group. Alkyl fragments dominate in the shadowed region with peak field intensities (10-15 V/nm), suggesting that the molecules undergo extensive fragmentation into short hydrocarbon species. Thus, the field-induced fragmentation process of CB molecules has been vividly depicted.

Complementary axial fragment maps (Fig. S8) further support this hierarchy, showing that more larger molecular fragments progressively transform into smaller ones as measurements proceed. This transformation may be related to i) more efficient cooling as tip shortens, or ii) a broadening of the shadowed zone due to the increased tip diameter. Together, these observations demonstrate the governing role of increasing field strength in intensifying fragmentation.

This analysis introduces a framework for controlling field-induced molecular fragmentation, particularly in soft organic systems where classical APT approaches of single atoms fall short. By revealing how subtle field gradients drive specific dissociation pathways, this approach provides a tool to probe failure mechanisms and degradation thresholds in organic functional materials operating under extreme electrical or interfacial conditions, such as in OLEDs, OFETs, or molecular layers under high-bias stress. It should also be noted that strategies for minimizing shadow-induced artefacts, particularly relevant for structure-focused APT studies, are provided in the Supporting Information (Fig. S9).

Building on the insights from fragmentation analysis, we next analyze molecular signals to distinguish between nematic and crystalline arrangements within LC systems, thereby demonstrating APT's structural resolving power. As first step, we assess whether the molecular distribution in a nematic mixture reflects homogeneity or yields any sign of segregation. We prepared a mixture of intermediate concentration, with nominally a 5CB molar fraction of 55 % to maximize the possible driving force to decomposition. For evaluation, we chose a region clearly dominated by molecular signals, containing over 200,000 molecules (Fig. 6a).

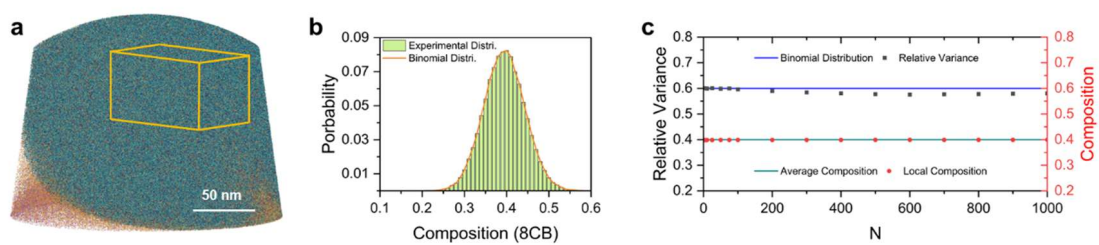

*Figure 6. a) Molecular mixing behavior in a $5CB_{(0.55)}8CB_{(0.45)}$ mixture studied within a reconstructed volume of $55 \times 96 \times 54$ nm³. b) Frequency distribution of 8CB molecules within spheres containing 100 molecules, with experimental histogram (lime) compared to an ideal distribution (orange curve). c) Local composition of 8CB molecules (red dots) compared with the average composition (green line), and relative variance (black dots) compared with the theoretical binomial distribution (blue line), analyzed across sphere sizes ranging from 1 to 1000 molecules.*

By centering on individual molecules and calculating the probability of finding 8CB among its nearest neighbors, the experimental histogram closely approximated a theoretical normal distribution (Fig. 6b). The local composition for the nearest $N$ = 1 to 1000 neighbors remained stable at ~0.4 (red dots in Fig.6c). This value is slightly below the nominal concentration (0.45) due to partial cleavage of 8CB into 5CB, which reduces the apparent 8CB count in the mass spectrum (Table S7). Nevertheless, the close overlap between the local compositions and the average composition (green line in Fig.6c) confirms the statistical homogeneity of the solution. Moreover, the concentration fluctuations (black dots in Fig. 6c) followed the expected relative variance ($\frac{\text{var}(c)}{c} = 1 - c$) of a binomial distribution (blue line in Fig. 6c), indicating no signs of nanoscale demixing[66,67] across all probed length scales. This statistical consistency, in agreement with phase diagrams and simulations[68–70] confirming a single nematic phase at room temperature. This indicates that the chosen cooling conditions are sufficient to retain their native structural mixed state at 318.5 K after vitrification, which establishes a solid basis for further studies of LC mixtures.

In contrast to a frozen homogeneous mixture, a pre-annealed pure 8CB sample reveals a markedly different picture. This sample was firstly heated to the isotropic phase at 318.15 K for deposition and then cooled to 283.15 K and kept at this temperature for 10 minutes before final quench freezing. As shown in Fig. 7a, a low-density area (blue or purple) appears at the center of the reconstructed volume of this sample, exhibiting a periodically layered structure in both molecular and fragmentation signals, while the surrounding high-density region (lime green or yellow) lacks this order. The Kernel density estimation revealed an average layer spacing of 0.731 nm (Fig. 7b), which is further confirmed by the clear peak in the Fourier transform (Fig. 7c). The periodic arrangement is very clear for the intact molecules and the larger fragments while less pronounced for the alkyl fragments and as expected vanishing for the hydroxides which mostly stem from chamber background. (Fig.7c).

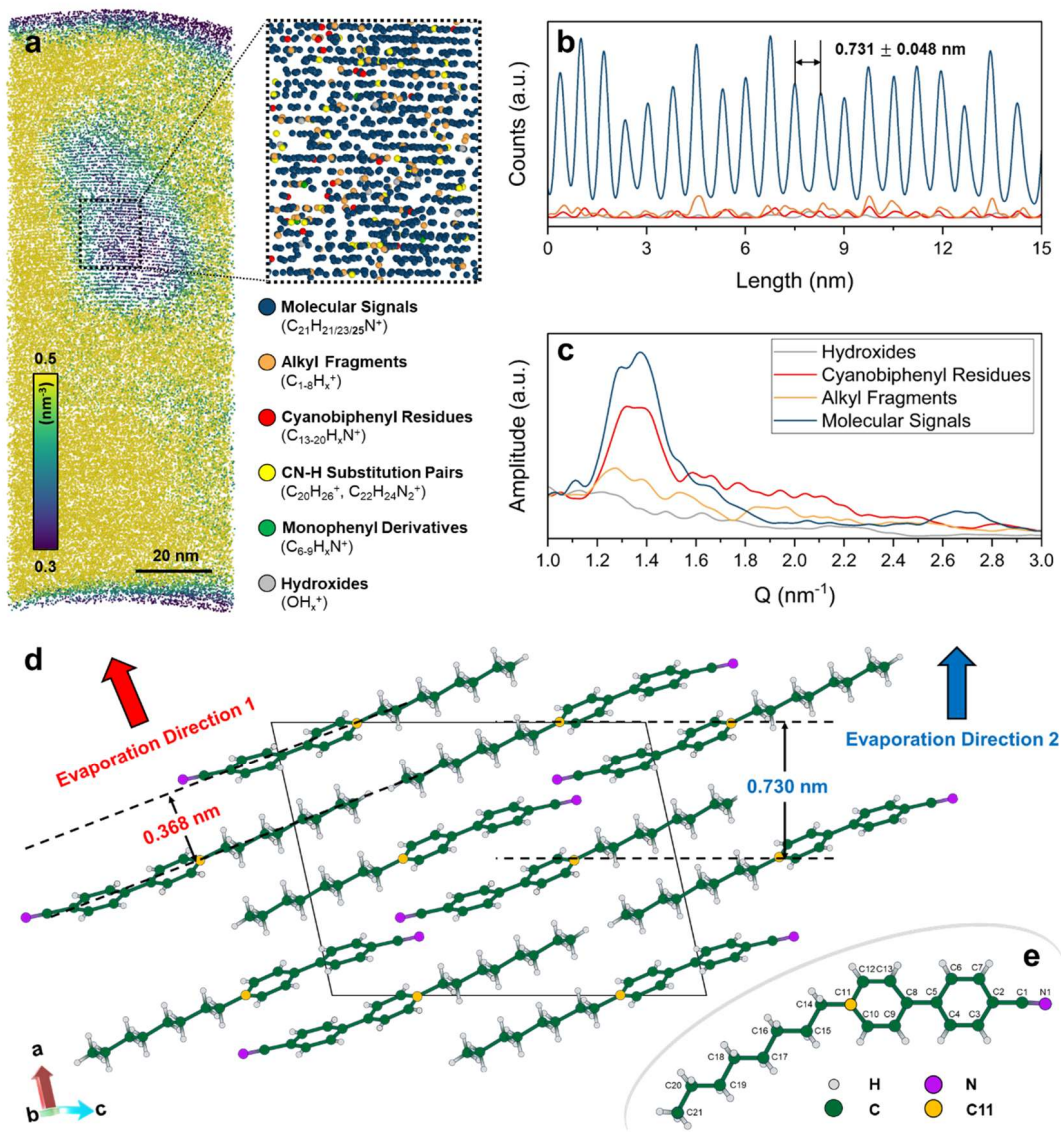

*Figure 7. a) Crystalline phase visualization through a local density map of a 10 nm-thick slice from the reconstructed 8CB measurement, highlighting a distinct layered structure in the low-density region. A zoomed-in view of a 17 nm-thick region in the Z direction is shown on the right, with major signals color-coded and labeled below. b) Lattice spacing determination using a kernel density estimation in spatial distribution analysis. c) Periodicity relationships between molecular and fragment signals illustrated via a fast Fourier transform (FFT) diagram. Hydroxide signals (grey), shown as a reference, exhibit no periodic correlation for contrast. d) Schematic representation of 8CB field evaporation directions relative to molecular orientation in the crystalline phase. e) Depiction of the atomic structure of 8CB, highlighting its molecular center through the designation of the C11 carbon atom (yellow).*

Since one-dimensional translational order is present, the phase must be either smectic or crystalline. However, the observed layer spacing is significantly smaller than the typical smectic phase spacing of 3.173 nm,[71] making a smectic phase unlikely. Furthermore, this sample was kept at 283.15 K for 10 minutes before quench-freezing, a temperature well below the crystalline transition temperature of 295 K.[72] This cooling duration is likely sufficient for nucleation of crystalline structures. Thus, we infer that the observed structure represents a crystalline region.

To assess the structural identity, we compared our measurements with existing crystallographic data. To date, only one stable crystalline phase of 8CB has been reported,[73] which has been identified by Kuribayashi and Hori as monoclinic *P*2$_1$/*n* (*Z*=4).[74] Figure 7d illustrates the corresponding unit cell viewing upon an *a-c* plane. In the $(40\bar{2})$ direction, 8CB molecules form a layered structure with the interlayer distance of 0.368 nm. However, this predicted spacing does not match with our observation. As a possible explanation that this spacing is not observed, one has to realize that in this orientation, the major electrical field would be aligned perpendicular to the molecular axis and so, the induced dipole moment is too weak to overcome the evaporation barrier.

In other crystallographic directions, a layered structure is less obvious. Since APT treats the entire molecule as a single evaporation entity, we need to approximate the molecular position. Inspired by previous studies on C$_{60}$ molecules,[75] we adopt the mass center as the reference point. However, unlike fullerene, the 8CB molecule is anisometric, requiring a more delicate selection of the center.

Among all atomic sites within 8CB, only C11 and C14 were found to reproduce a well-defined layered structure when aligned along specific crystallographic directions. We choose C11 as the reference point because it is located in the rigid aromatic core of the 8CB molecule (Fig. 6e), making it structurally stable and reproducible during APT evaporation. In contrast, C14, which could also be considered as a center atom, is part of the flexible alkyl chain. Thus, it is more susceptible to conformational changes, potentially compromising reconstruction accuracy. By simplifying the 8CB molecule as an equivalent point at central C11 atom (yellow atoms in Fig. 7d), we deduce that the layer spacing along the (200) crystallographic direction is 0.730 nm, which aligns very well with the experimental results. In this inclined direction a more significant dipole moment is induced that obviously can trigger field evaporation.

The identification of a crystalline domain embedded within a low-density region, in clear spatial contrast to the surrounding high-density amorphous matrix, offers the rare opportunity to visualize the solid-liquid interface in an organic molecular system. This observation warrants further investigations, especially in mixed systems where segregation to the interface becomes an interesting feature.

## 3. Conclusion and Perspectives

This study presents the first application of cryo-APT to thermotropic LC systems, bridging two previously disconnected domains through their intrinsic properties. The high chemical sensitivity and mass resolution of APT enabled detection of low-abundance fragment signals, whose anisotropic spatial distribution reveals a laser-induced shadow effect amplified by LC fluidity. By reconstructing the tip geometry and computing local electric fields, we quantified how field gradients govern CB molecular dissociation pathways. Meanwhile, the high spatial resolution of APT allowed direct visualization of molecular lattice planes in crystalline 8CB, highlighting its capability to capture long-range order in frozen LC systems.

More broadly, our findings establish cryo-APT as a powerful analytical platform for soft matter research, capable of simultaneously preserving molecular integrity and liquid-phase structural features, which is a dual achievement rarely attainable due to the combined challenges of radiation sensitivity in organic molecules and thermodynamic metastability in liquids. The >90% retention of intact molecules, along with clear resolution of nematic miscibility in mixture and crystalline layering in pure 8CB, demonstrates APT's applicability to chemically complex and structurally heterogeneous materials beyond LCs. At the same time, thermotropic LCs, with their well-characterized ordering, offer a benchmark for validating spatial fidelity, underscoring the robustness of our cryo-APT protocols for broader application.

While the present study does not yet resolve molecular conformations or orientational states, the observed controllable dissociation behavior of nCB molecules suggests that tailored fragmentation strategies may enable future spatial mapping of dipolar alignment or director fields. Such developments could open new avenues for addressing long-standing challenges in LC research, including the influence of chiral dopants[76,77] and other additives,[78,79] the detection of nanosegregation[70,80] and cybotactic clusters,[81] the characterization of anchoring at interfaces or under confinement,[82,83] and the structural elucidation of emerging ferro- and antiferroelectric nematic phases.[84] Collectively, these capabilities position cryo-APT as a transformative tool for resolving molecular ordering in soft matter systems at unprecedented resolution.

## 4. Experimental
### 4.1 Materials
5CB and 8CB (purity: 98%) were purchased from Merck and used without further purification. Both compounds appeared opaque in the liquid state at ambient temperature. For mixture preparation, each component was individually heated to 318.15 K to reach the isotropic phase, followed by combining the two in desired ratios using a vortex mixer.

Tungsten wires (Alfa Aesar) with diameters of 75, 50, and 30 µm were tested. Among these, 50 µm was the minimum diameter at which the wire remained straight after cryogenic fracture, producing the smallest stable platform for sample deposition.

### 4.2 Transfer, Freezing Etching and Shaping
Frozen sample transfer was performed using a Leica VCT500 transfer system at 89 K under a vacuum better than 0.02 mbar. Freeze-etching was conducted in a Leica EM ACE600 high-vacuum coater at 185 K for 60 min at $10^{-5}$ mbar to remove frozen water contaminant from the surface. Sample preparation within the dual beam microscope (Thermofisher SCIOS) was performed on a custom-made cryo stage connected by two copper bands connected to a liquid nitrogen dewar, maintaining the temperature close to 123 K.

### 4.3 Measurement Conditions
The prepared tip sample was transferred through a cryo-transfer port into a custom-made atom probe,[85] featuring a 120 mm delay line detector system and an open area ratio (OAR) of 50 %. Tips samples were measured under an operating Clark MXR laser system at a wavelength of 355 nm, with a 250 fs pulse length and a 30-micron spot size. All nine tip samples (Fig. S3) were measured at 60 K, with the laser intensity of 0.127 nJ/nm$^2$ per pulse at 100 kHz. The evaporation rate was controlled between 400 and 800 entities per second.

### 4.4 Data Process
The measured datasets were processed using custom-developed Python modules, which are freely available online.[86] The processing workflow included mass spectrum calibration, chemical identification through analytical fitting of the spectra, and final three-dimensional reconstruction. Reconstruction was performed using a taper geometry, with the shaft angle and initial tip radius determined from SEM images acquired during sample preparation. The volume reconstruction followed the original projection protocol described by Bas *et al*.,[58] incorporating the geometric boundary conditions established by Jeske *et al*..[87] Surface profile reconstruction was conducted using the algorithm proposed by Beine and Schmitz.[88] Visualization of the reconstructed data was performed using OVITO.[89]

### 4.5 Calculation of the electrical field
To evaluate the field distribution at a field emitter with an arbitrary, but convex surface shape, the surface profile was discretized (10.000 points) and Robin's method was applied to calculate the charge distribution. Basically, each discrete point of the relevant surface was initially loaded with a unit charge and subsequently, these charges were redistributed among the surface points by iterative application of

$$\frac{q_l^{(i+1)}}{s_l} = \frac{1}{2\pi} \sum_{k \neq l} q_k^{(i)} \frac{\mathbf{n}_l \cdot \mathbf{r}_{l,k}}{|r_{l,k}|^3}$$

until a stable end distribution was achieved, see [65]. ($q_l^{(i)}$, $s_l$, $\mathbf{n}_l$ and $\mathbf{r}_{l,k}$ denote the charge at surface point $l$ in iteration step $i$, the area of the surface segment $l$, the outwards normal to this segment and the distance vector between points $l$ and $k$, respectively). Performed iterations must warrant that total charge remains constant. The local field strength at the tip surface is calculated from the finally achieved charge distribution as:

$$F_l = \frac{q_l}{\varepsilon_0 \, s_l}$$

In the described form of the algorithm, the absolute value of the field strength is arbitrary, only the relative variation along the surface is meaningful. We calibrated the high field region to 10 V/nm, since such evaporation threshold is typically required to split alkyl chains into small $C_xH_y$ hydrocarbon fragments[27] which are observed in the measurement.

**Acknowledgments:**
We gratefully acknowledge financial support from the Deutsche Forschungsgemeinschaft (DFG, German Research Foundation) under Project-ID 358283783 - SFB 1333. The authors thank B. Gault for inspiring discussions that motivated the fragmentation analysis in this work.

**Supporting Information Available:**
Additional figures showing electron and ion beam effects on LC droplets (Figure S1), SEM images of nine successfully prepared APT specimens (Figure S2), voltage histories and reproducibility of spectra (Figures S3–S4), detailed mass spectrum peak assignments for 5CB, 8CB, and mixtures (Tables S2–S4), schematic illustrations of fragmentation behavior (Figures S5–S6), stoichiometry calculations (Figures S7–S8, Tables S5–S7), and analysis of laser-induced shadow effects and mitigation strategies (Figure S9). References for Supporting Information are also provided.

**Declaration of generative AI and AI-assisted technologies in the manuscript preparation process:**
During the preparation of this work, the author used ChatGPT (OpenAI, San Francisco, CA, USA) to improve the language and readability of the manuscript. After using this tool, the author thoroughly reviewed, edited, and integrated the content to ensure coherence and accuracy. The tool was not used for data analysis, result interpretation, or any other part of the scientific process. The author takes full responsibility for the content of the published article.

89. Stukowski, A. Visualization and analysis of atomistic simulation data with OVITO–the Open Visualization Tool. *Model. Simul. Mater. Sci. Eng.* **18**, 015012 (2010).
30